\documentclass{mecbic}
\providecommand{\anno}{2011} \providecommand{\firstpage}{3}
\providecommand{\lastpage}{4}
\usepackage{breakurl}

\usepackage{color}
\usepackage{amsmath}
\usepackage{amssymb}
\usepackage{stmaryrd}
\usepackage{url}
\usepackage{ifpdf}
\usepackage{fancyvrb}
\usepackage{upgreek}
\usepackage{mathpartir}
\usepackage{wasysym}
\usepackage{graphicx}

\newcommand{\ccs}{\textsc{ccs}}
\newcommand{\rccs}{\textsc{rccs}}
\newcommand{\DNA}{\textsc{dna}}

\title{Reversibility in Massive Concurrent Systems}

\author{Luca Cardelli \institute{Microsoft
Research, Cambridge} \email{luca@microsoft.com} \and Cosimo Laneve \institute{Universit\`a di
Bologna}\email{laneve@cs.unibo.it}}

\def\titlerunning{Reversibility in Massive Concurrent Systems}
\def\authorrunning{Luca Cardelli and Cosimo Laneve}

\begin{document}
\maketitle
\pagestyle{plain}
\pagenumbering{arabic}
\setcounter{page}{3}

\begin{abstract}

\end{abstract}

Reversing a (forward) computation history
means undoing the history.
In concurrent systems, undoing the history is not performed in a deterministic way but in a
causally consistent fashion, where states that are reached during a
backward computation are states that could have been reached during the
computation history by just performing independent actions in a different
order. In {\rccs}, Danos and Krivine achieve this
by attaching a memory $m$ to each process $P$, in
the monitored process construct~$m : P$. Memories in {\rccs} are
stacks of information needed for processes to backtrack.
Alternatively, Phillips and Ulidowski propose a technique for reversing process calculi
without using memories. In this technique,
the structure of processes is not destroyed and the progress is noted
by underlining the actions that have been performed. In order to tag the communicating processes, they generate unique identifiers on-the-fly during the communications.

These foundational studies of reversible and concurrent computations
have been largely stimulated by areas such as chemical
and biological systems 
-- called {\it massive concurrent systems} in the following -- where operations are reversible, and only an appropriate injection of
energy and/or a change of entropy can
move the computational system in a desired direction.

However there is a mismatch between chemical and biological
systems and
the above concurrent formalisms. In the latter ones, reversibility
means {\it desynchronizing processes that actually
interacted in the past} while, in massive concurrent systems,
reversibility means {\it reversibility of
 configurations}. In order
to make massive concurrent systems reversible with the process
calculus meaning,
one has to remember the position and
momentum of each molecule, which is precisely contrary to the well-mixing
assumption of biochemical soups, namely that the probability of collision
between two molecules is independent of their position ({\it cf.} Gillespie's algorithm).

To comply with the well-mixing assumption,
notions of causality and independence of events need to be
adapted to reflect the fundamental fact that different processes of the
same  species are indistinguishable. Their interactions can cause
effects, but not to the point of being able to identify the precise
molecule that caused an effect.
We introduce an algebra for massive concurrent systems, called
{\it reversible structures}, and,
following L\'evy, we define an equivalence on computations
that abstracts away from the order of causally independent reductions
-- the {\it permutation equivalence}. Because of multiplicities
this abstraction
does not always exchange
independent reductions. For example, two reductions that use a same
signal cannot be exchanged because one cannot grasp
whether the two reductions are competing on a same signal or are using
two different occurrences of it.
Notwithstanding this inadequacy, permutation equivalence in
reversible structures yields a standardization theorem
that allows one to remove converse reductions from computations.

Reversible structures may implement significant {\ccs}-style interaction
patterns (Cardelli already noticed this by studying a class of reversible systems -- the
{\DNA} chemical systems).
Consider for example a binary operator that takes two input molecules and produces one unrelated output molecule when (and only when) both inputs are present. It is too difficult to engineer the input machinery
in order to any possible pattern of interaction, and to produce the output molecule out of their own structure. This operator is therefore implemented by an artifact that binds the two inputs one after the other
and then releases the output out of its own structure.
Of course, if the second input never comes it must release the first input, because the first input may be legitimately used by some other operator. This means that the binding of the first input must be reversible, and the natural reversibility of reversible structures is exploited to achieve the correctness.

In order to bridge the gap between reversible process calculi and
massive concurrent systems, we consider reversible structures where multiplicities are dropped (terms have multiplicity one) -- the
{\it coherence}
constraint. Coherence in this strong sense is not realizable in
well-mixed chemical solutions, but may become realizable in
the future if we learn how to control individual
molecules. We demonstrate that coherent reversible structures
implement the {\it asynchronous} fragment of {\rccs}.

The exact distance between coherent and uncoherent
reversible structures (that is, between reversible process calculi and
massive systems) is manifested by the computational complexity of the
reachability problem (verifying whether a configuration is reachable from
an initial one).
We demonstrate that reachability in coherent
reversible structures has a
computational complexity that is quadratic with respect to the size of
the structures, a problem that is {\sc expspace}-complete in generic structures.

Our study prompts a thorough analysis of reversible
calculi where processes have multiplicities and the causal dependencies
between copies may be exchanged. Open questions are (i)
What synchronization schemas can be  programmed in massive concurrent
systems? (ii) Are there other constraints, different than coherence,
such that relevant bio-chemical properties retains better algorithms
than in standard structures? (iii) What is the theory of
 massive (reversible)
systems {\it with irreversible operators} and what is the relationship with
standard programming languages?

\end{document}